# Transformation of Wiktionary entry structure into tables and relations in a relational database schema [1]

Andrew Krizhanovsky

Institution of the Russian Academy of Sciences St.Petersburg Institute for Informatics and Automation RAS
V.0. 39, 14-th Line, St.-Petersburg, Russia, 199178, SPIIRAS
Phone +7 (812) 328-80-71

andrew dot krizhanovsky@gmail.com

ABSTRACT

This paper addresses the question of automatic data extraction from the Wiktionary, which is a multilingual and multifunctional dictionary. Wiktionary is a collaborative project working on the same principles as the Wikipedia. The Wiktionary entry is a plain text from the text processing point of view. Wiktionary guidelines prescribe the entry layout and rules, which should be followed by editors of the dictionary. The presence of the structure of a Wiktionary article and formatting rules allows looking at the entry from the object-oriented point of view. Then, (1) the article, its sections and subsections would correspond to classes, (2) the existence of a subsection in the section indicates that there is a relation between the corresponding class "section" and class "subsection". This correspondence makes it possible to transform the flat text of Wiktionary article into the object-oriented form, i.e. to create instances of objects, to fill data fields of the objects on the basis of Wiktionary data. The consequent result is a creation of an application programming interface in order to communicate with objects, and, in fact, to deal with data extracted from the Wiktionary. On the other hand, in order to provide a quick and easy data treatment, it requires the data to be stored in the database. So, the paper describes the task of transformation of the Wiktionary entry structure into tables and relations in a relational database schema, which is a part of a machine-readable dictionary (MRD). In other words, it describes how the flat text of the Wiktionary entry was extracted, converted, and stored in the specially designed relational database. The MRD contains the definitions, semantic relations, and translations extracted from the English and Russian Wiktionaries. The parser software is released under the open source license agreement (GPL), to facilitate its dissemination, modification and upgrades, to draw researchers and programmers into parsing other Wiktionaries, not only Russian and English.

Keywords: Wiktionary, Dictionary, Thesaurus, Lexicography, Machine-readable dictionary, Parser.

## 1. INTRODUCTION

Wiktionary is a unique resource and it could be useful for a wide variety of NLP tasks. But it cannot be used directly. There is a need to develop a software tool which makes it possible to convert the Wiktionary articles into a more suitable form for computer manipulation and processing, such as machine-readable dictionary (MRD).

The problem is that, from the programmer's point of view, when a computer program tries to deal with the entry directly, the Wiktionary entry is an ordinary text. The structure of the entry is described in the Wiktionary guidelines, but it exists only in the head of the editor, and there are no any special techniques to control the user input and to prevent an erroneous or malicious action. In the absence of such constraints, dictionary's editors may conduct experiments for the purpose of designing new, better formatting rules. Nevertheless, this flexibility of the format turns out to be a complex problem when there is a need to develop the software (parser) to extract data from the Wiktionary. It is possible that the development of such parser resembles a development of a modern browser, which should provide correct treatment of any erroneous HTML pages, at the same time it should present the maximum of the recognized information from the HTML page.

The specific of the Wiktionary is that it is created by the community of enthusiasts, and it is probably that not all of them are professional lexicographers. The structure of a dictionary entry is gradually, but constantly changing, since community experts regularly discuss and work out new and better rules. Also it should be taken into account that Wiktionary is permanently growing in number of entries and in the scope of languages. Now the English Wiktionary contains about 760 different languages, and the parser recognizes 169 language codes; in the Russian Wiktionary

[1] See Russian version of this paper: http://scipeople.com/publication/100231/

there are 343 languages, and the parser recognizes 337 language codes.

In the previous work [5] the problem of data extraction from the Wiktionary was considered from the viewpoint of the computer program development, namely, requirements to the parser and the program architecture. In this paper more attention will be devoted to the data organization, that is the relationship between the Wiktionary entry layout and the structure of the database of the machine-readable dictionary will be considered.

Section 2 presents some background work on the research and software related to the Wikipedia. This is followed by a review of the related work on the machine-readable dictionaries construction, the building of thesauri, the data retrieval from the Wiktionary and the Wikipedia.

Section 3 describes the relationship between the Wiktionary entry layout and the structure of the database of the machine-readable dictionary. The discussion concludes the paper.

## 2. BACKGROUND AND RELATED WORK

Before the creation of the Wiktionary parser, the open-source software for the Wikipedia data extraction, the wikitext parsing, the indexing of Wikipedia texts [7], the search for related terms by analyzing Wikipedia internal links structure [8] was developed in our lab. This software can be used to create a parser, when the following conditions are met:

- Software must be written in the Java programming language.
- Wiktionary dump should be passed to the MySQL database.

Undoubtedly, the task of creating of machine-readable dictionaries existed long before Wiktionaries come on the scene [9], [15]. However this amazing lexicographic resource, which is the Wiktionary, appeared only now.

Manually created thesauri, e.g. WordNet, have enjoyed considerable popularity in natural language processing. Thesauri which are filled automatically with data extracted from Wikipedia or from the Web are also actively used.

Not only the Wiktionary, but also the Wikipedia could be considered as a thesaurus. There are special algorithms to extract semantic relations from Wikipedia. E.g. the hyponyms and hypernyms are extracted from the Japan Wikipedia [13]. More types of semantic relations were retrieved from the English Wikipedia in order to construct an ontology [4].

Nevertheless, there is a small number of studies related directly to the Wiktionary. A rare example is the paper [16] which describes application programming interfaces for Wikipedia and Wiktionary (English and German Wiktionaries).

There is a number of works devoted to the comparison of Wiktionaries with other thesauri. In our previous research [6] the related terms search based on the Russian Wiktionary was compared to WordNet based algorithms. The WordNet won.

The comparative study of the three resources German Wiktionary, GermaNet and OpenThesaurus that analyzes both topological and content related properties is presented in the paper [10]. It was revealed that the German Wiktionary contains the lowest number of semantic relations (157 thousands, June 2009).

The Wiktionary in turn can be used to construct other thesauri. Thus, for example, French and Slovene WordNet were built by data extracted from different resources including French, Slovene, and English Wiktionary [2].

## 3. RELATIONSHIP BETWEEN THE WIKTIONARY ENTRY LAYOUT AND THE STRUCTURE OF THE DATABASE OF THE MACHINE-READABLE DICTIONARY

The structure of the Wiktionary entry is defined clearly, rigorously and strictly by rules. These rules are presented in the English Wiktionary[2], in the Russian Wiktionary[3], and it seems that they are presented in other 168 wiktionaries. [4]

It should be noted that the Wiktionary (in the same way as the Wikipedia) is supported by MediaWiki software. But MediaWiki don't take into account this structure of the Wiktionary entry. That is, in the MediaWiki database there are, roughly speaking, only two fields: a name of an entry and a wiki-markup text of an article.

Since, there are an entry layout and formatting rules, it follows that the Wiktionary entry could be considered as a very interesting object from the automatic data extraction point of view, e.g. with the help of regular expressions [3]. This automatic data extraction will allow a transforming of an implicit structure (i.e. structure that is understandable for the Wiktionary reader) into the explicit, unambiguous, machine-understandable structure. After that, the Wiktionary data could be easily used in projects related to the language processing.

---

[2] See http://en.wiktionary.org/wiki/Wiktionary:ELE

[3] See http://ru.wiktionary.org/wiki/Викисловарь:Правила_оформления_статей

[4] See http://meta.wikimedia.org/wiki/Wiktionary/Table

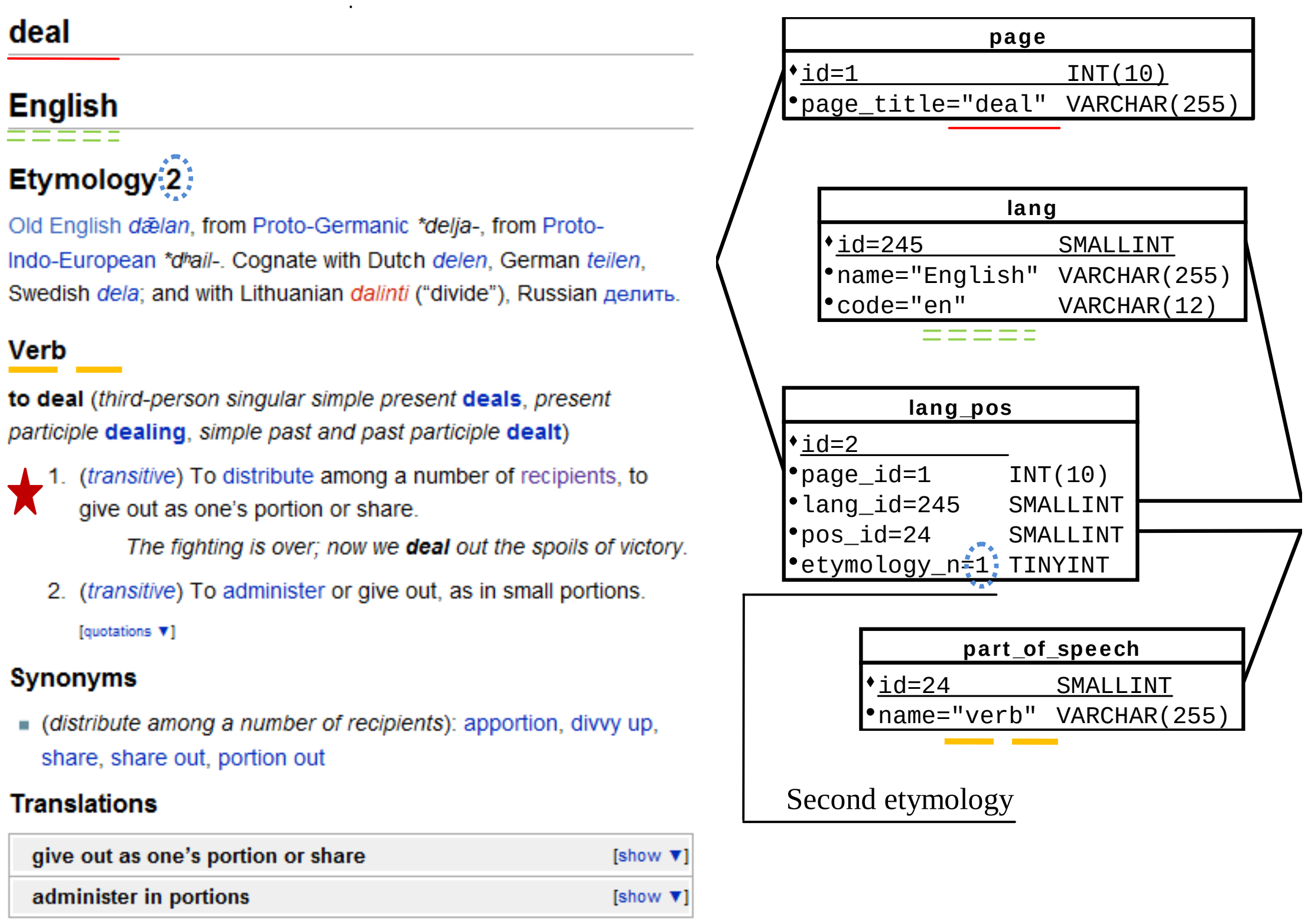

**Figure 1. The mapping of a part of the entry *deal* from the English Wiktionary (left) to fields of tables of the machine-readable dictionary: the title of the article (*deal*), the language (*English*), the part of speech (*verb*)**

Below, the fragments of a Wiktionary entry will be presented. Tables of the machine-readable dictionary (required in order to store data extracted from the Wiktionary) will be shown.

For the reader's convenience, only a part of the English Wiktionary entry *deal*[5] is presented in Figure 1.

With the help of dotted and continuous lines, it is shown in Figure 1 that there is a correspondence between each element of the entry (left) and fields of the tables (right). The structure of the machine-readable dictionary corresponds to the structure of the source data (i.e. to the Wiktionary entry layout) as much as possible. There are the following interrelated tables in the machine-readable dictionary (at the right in Figure 1):

- *page* – it is a key table in the database. It contains a unique identifier (the field *id*) and a name of an entry (the field *page_title*). The name is underlined by a continuous line.
- *lang_pos* – it is a second main table. It is a pointer to the part of the entry precisely identified by three parameters: the language of this part of the entry (field *lang_id*), the name of the part of speech (field *pos_id*), and the etymology number (field *etymology_n*). The number of etymology (dotted circle in Figure 1) allows enumerating homonyms. An ordinal number of the second homonym (*Etymology 2*) is shown in Figure 1, and in the database the meaning of the field *etymology_n* is equal to 1 (counting from zero).
- *lang* – it binds a unique identifier to a language name (e.g. *Swedish*) and a language code (e.g., *sw*). There is a full list of languages names and codes that are used in the English Wiktionary.[6] Now the parser recognizes 169 languages in the English Wiktionary[7] and 337 languages and codes in the Russian Wiktionary.[8] An example in Figure 1 describes the English entry word with language

[5] See the full entry http://en.wiktionary.org/wiki/deal

[6] See http://en.wiktionary.org/wiki/Wiktionary:Index_to_templates/languages

[7] See http://en.wiktionary.org/wiki/User:AKA_MBG/Statistics:Translations

[8] See http://ru.wiktionary.org/wiki/Участник:AKA MBG/Статистика:Переводы

code *en*, where language name and code are underlined by the double-dashed lines.

- *part_of_speech* – the table binds a name of a part of speech (the field *name*) to an identifier (the field *id*), where identifier is used in the table *lang_pos* (the field *pos_id*). An example in Figure 1 describes a *verb*, where the name of the part of speech is underlined by the dashed line.

The first meaning in Figure 1 is marked by the star. It will be discussed in more detail further on.

An example of formatting of a word's definition in the English Wiktionary is shown in Figure 2: from a Wiktionary editor's point of view (*wiki markup*) and from a reader's point of view (*result*). The interrelated tables (in the database of the machine-readable dictionary) related to this definition are presented in Figure 2 (bottom).

The developed parser extracts semantic relations and translations that are stored in the Wiktionary entry. But in the entry the translations and semantic relations are not bound (i) to the whole entry (it corresponds to the table *page*), or (ii) to entry part which corresponds to a language and a part of speech (the table *lang_pos*).

In the entry, hence in the machine-readable dictionary, the translations and semantic relations are bound to the particular meaning of the word. As example, the first meaning is considered in Figure 2. This meaning is marked by the star and the corresponding wikitext is underlined by the dashed line.

Thus, it was needed to add the table *meaning* to the machine-readable dictionary. The table contains the following fields:

- the field *lang_pos_id* points to a part of speech, a language name and a name of an entry via the table *lang_pos*;
- the field *meaning_n* defines a meaning number (counting from zero);
- the field *wiki_text_id* provides an access to the text of the definition, which is stored in the table *wiki_text* (it is underlined by the dashed line).

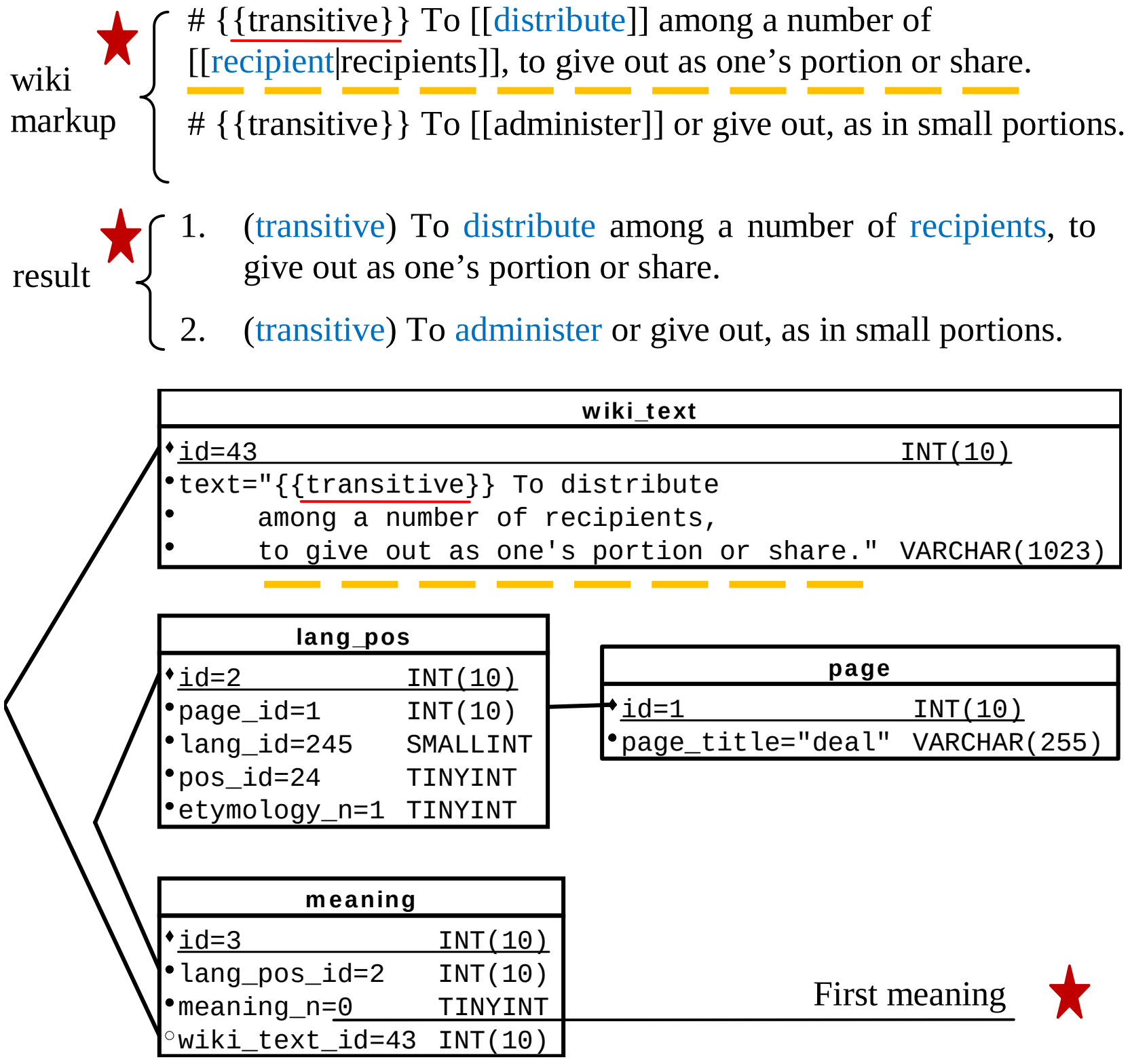

**Figure 2. The first meaning of the English verb *deal* presented in tables of the machine-readable dictionary**

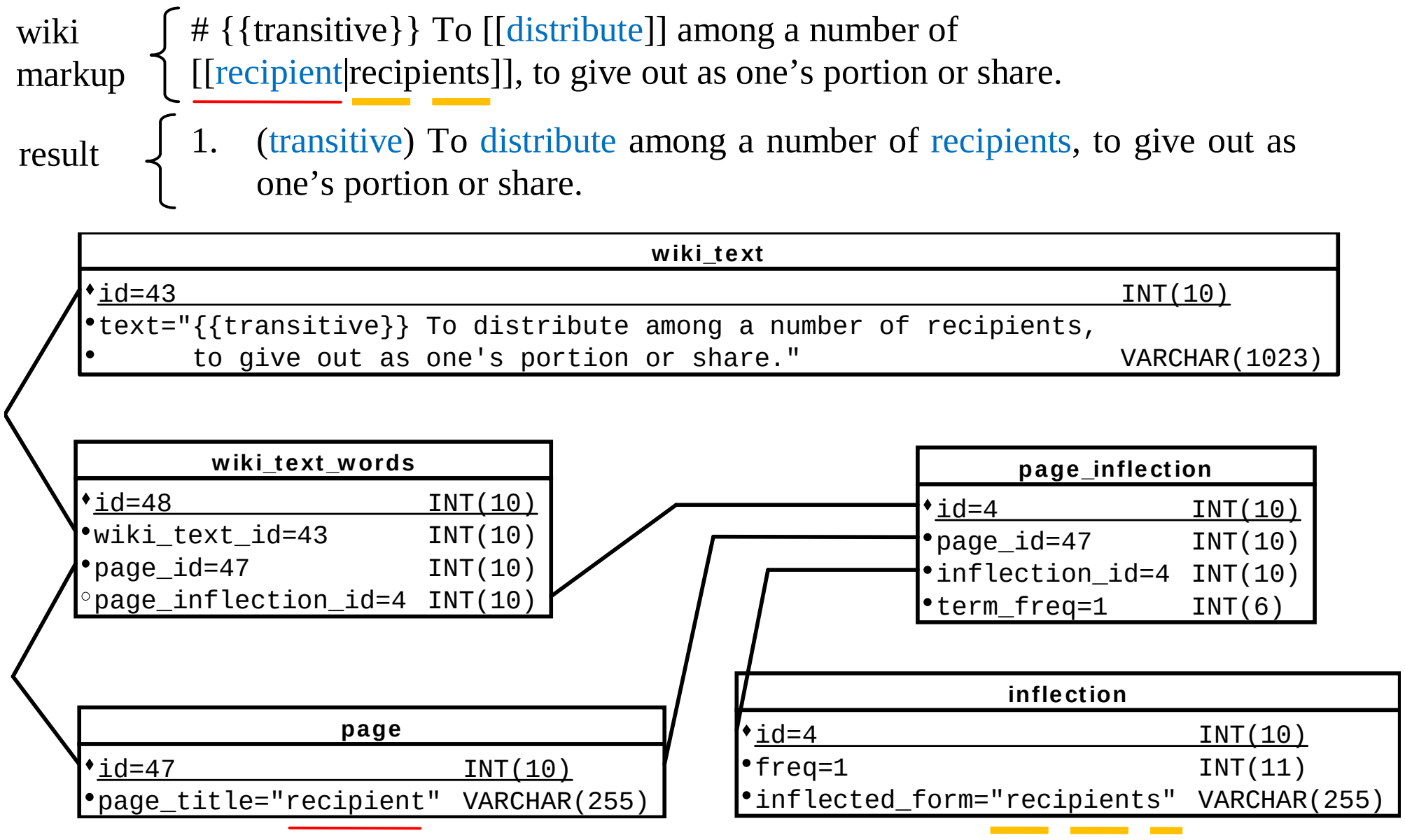

**Figure 3. Internal links (denoted by square brackets in a wikitext) are stored in the tables *wiki_text_words*, *page_inflection*, and *inflection***

Internal link (wikilink or free link) links a page to another page within a wiki website. In a case of the Wiktionary, internal link links to a lemma (the canonical form of an inflected word). For example, *[[recipient|recipients]]*, shown in Figure 3, displays as *recipients*, but links to the *recipient* wiki page.

During the development of the parser it was decided to store this information in the database of the machine-readable dictionary because it could be used for different purposes, e.g. in order:

- *To find lemma forms for the words*. Someday the Wiktionary will contain all words of all languages. Then information about lexeme (all the inflected forms of a term) will be presented in the entry. But now, when the Wiktionary starts to exist, there are a lot of red links that hint about missing Wiktionary entries. Thus, for some words an internal link is the only source of information about lemma forms yet.
- *To find the most frequent forms of the word*.
- *To find a list of keywords (or semantically related words) that represents the meaning of the entry* (with the help of the thesaurus found in the Wiktionary).

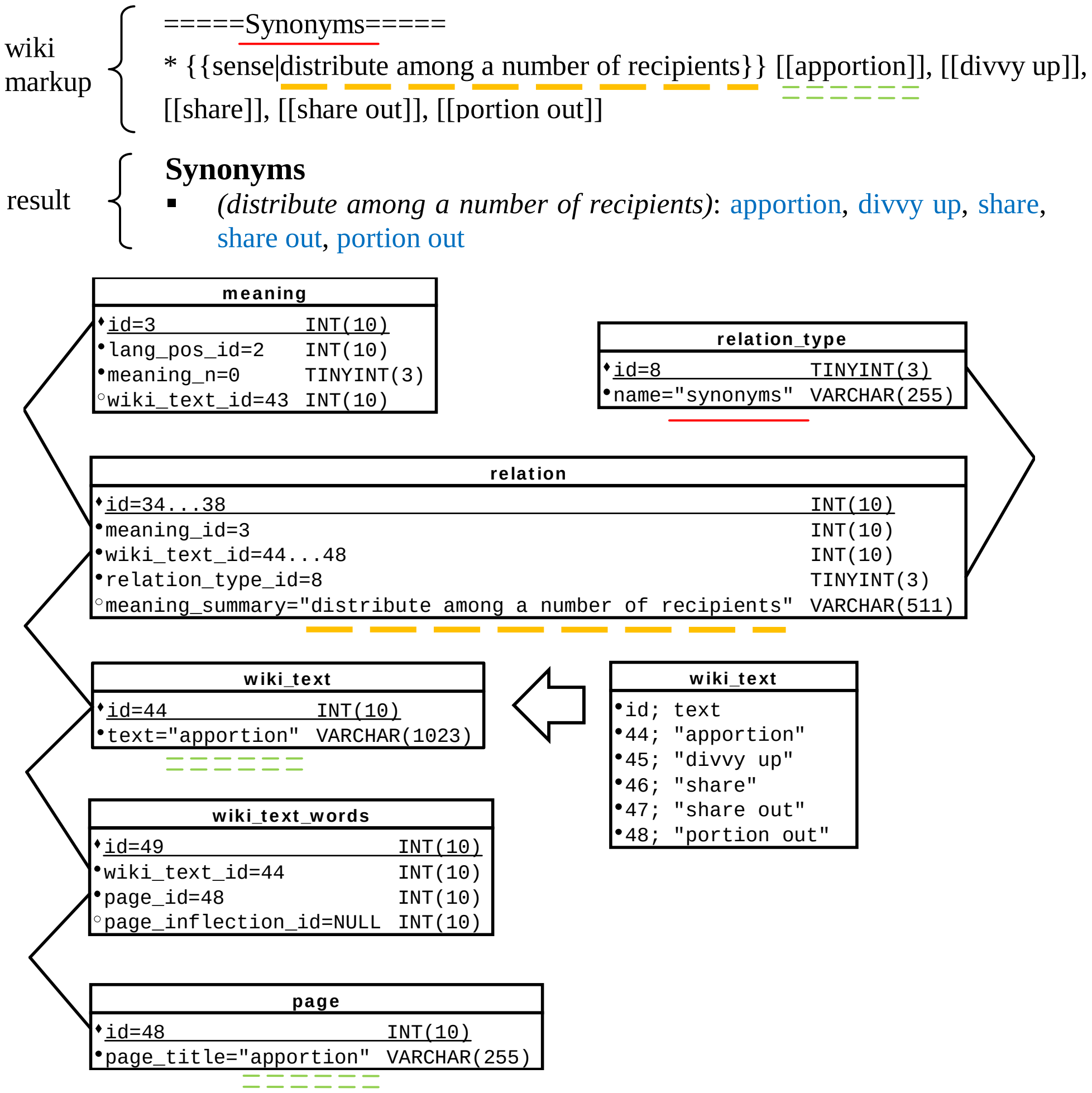

**Figure 4. The list of synonyms (*apportion, divvy up, share, share out, portion out*) of the first meaning (meaning_n=0) of the verb *deal* stored in the machine-readable dictionary database**

In order to store semantic relations to the machine-readable dictionary, two tables (*relation* and *relation_type*) can be used (Figure 4). The table *relation_type* is filled only once (before the Wiktionary parsing begins); it contains 9 types of semantic relations presented in the English Wiktionary.[9]

The parser extracts a definition summary (which presents in a condensed form a summary for a list of synonyms) from the template *{{sense|}}* (the summary is underlined by the dashed line in Figure 4). A text of the summary is stored to the field *meaning_summary* of the table *relation*.

As has been shown in Figure 4, the fields *id* and *wiki_text_id* in the table *relation* have ranges of values. This emphasizes that a list of synonyms (stored in the table *wiki_text*) is bound to the given meaning of the word (i.e. to the table *meaning*) via the table *relation*.

Figure 4 shows that the storing of semantic relations produces a duplication of information in the tables *wiki_text* and *page* (e.g. the synonym *apportion* is stored in both tables, it is underlined by double-dashed lines). It is a particular case of data redundancy, when a wikitext consists of only one word which is a hyperlink (more precisely, internal link) linked to some Wiktionary entry. In the more common case (see previous Figure 3), the wikitext is a phrase, and there are no any Wiktionary entries corresponding to this phrase, i.e. there is no redundancy.

---

[9] See http://en.wiktionary.org/wiki/Wiktionary:Semantic_relations

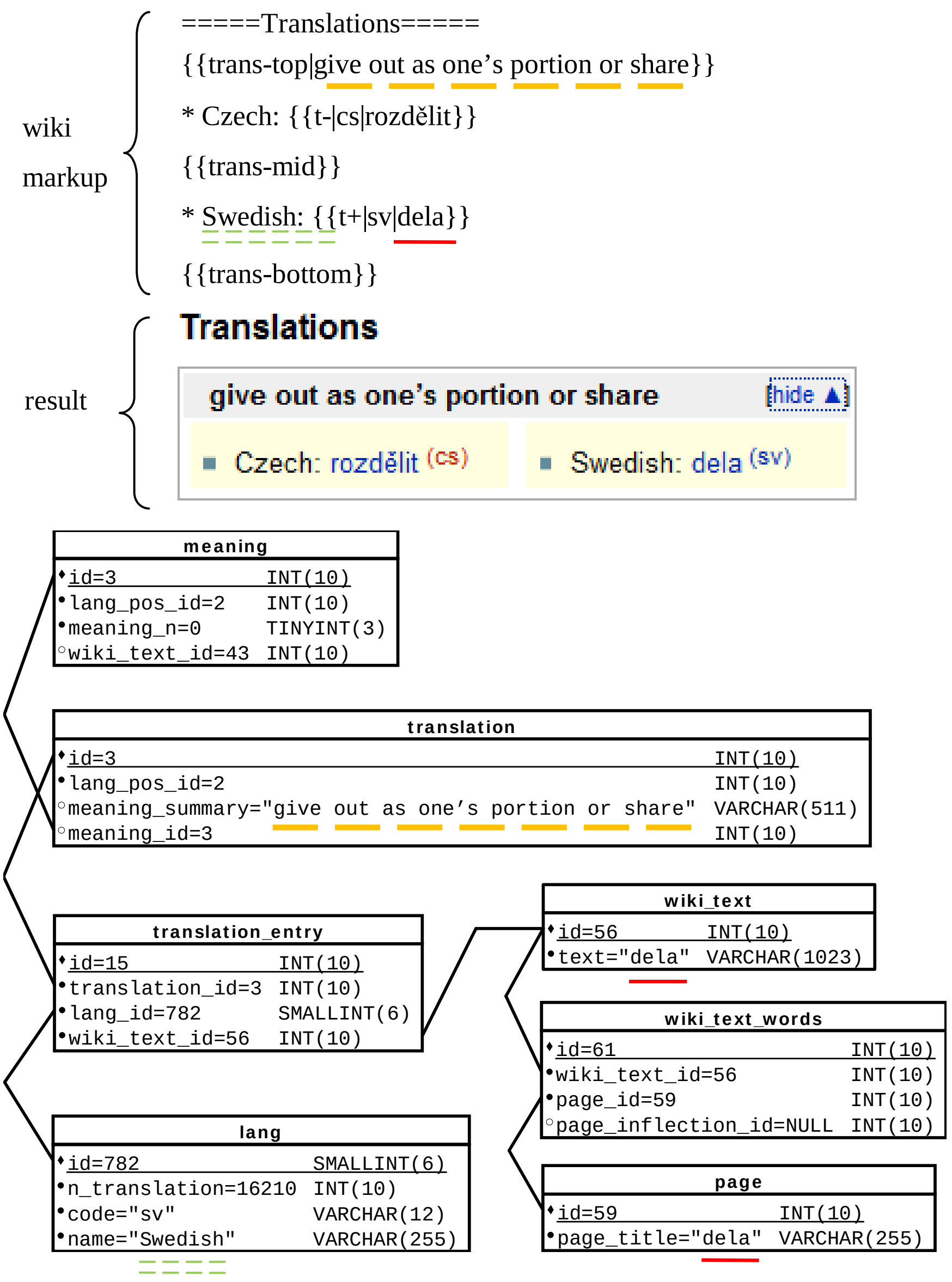

**Figure 5. The section *Translation* of the entry *deal* stored in tables of the machine-readable dictionary database**

An example of formatting of a word's translation in the English Wiktionary is shown in Figure 5 (top). The translation of the first meaning of the verb *deal* to Sweden stored in the machine-readable dictionary database is presented in Figure 5 (bottom). The translation *dela* is underlined by a continuous line; the language name is underlined by the double-dashed lines. The code of Sweden language is *sv*, it is stored in the table *lang*. (with the record number of 782).

The field *n_translation* in the table *lang* (Figure 5) indicates a number of translations from the main language (the English in the English Wiktionary) to the target language. That is, there are 16210 translations to Sweden, in accordance with the data extracted from the English Wiktionary as of August 24, 2010.

The section *Translation* corresponds to the following interrelated tables in the MRD database:

- For each definition (the record in the table *meaning*) there is one record in the table *translation* (the relation one-to-one). If there is a definition summary (as a parameter of the template *{{trans_top|}}*), then this summary will be written

to the field *translation.meaning_summary*, it is underlined by the dashed line (Figure 5).

- One record in the table *translation* corresponds to many records in the table *translation_entry*, one record for one language and one translation (the relation one-to-many). There are two translations in Figure 5 (top), one translation for one language, namely, translation to Czech (*rozdělit*) and to Sweden (*dela*). Thus, in the table *translation_entry* will be exactly two records with links to:
    - o a target translation text (the field *wiki_text_id*);
    - o a target translation language (the field *lang_id*);
    - o a particular meaning (*meaning.id*) via the table *translation* (the field *translation_id*).

The whole picture of the MRD database is presented in Figure 6.

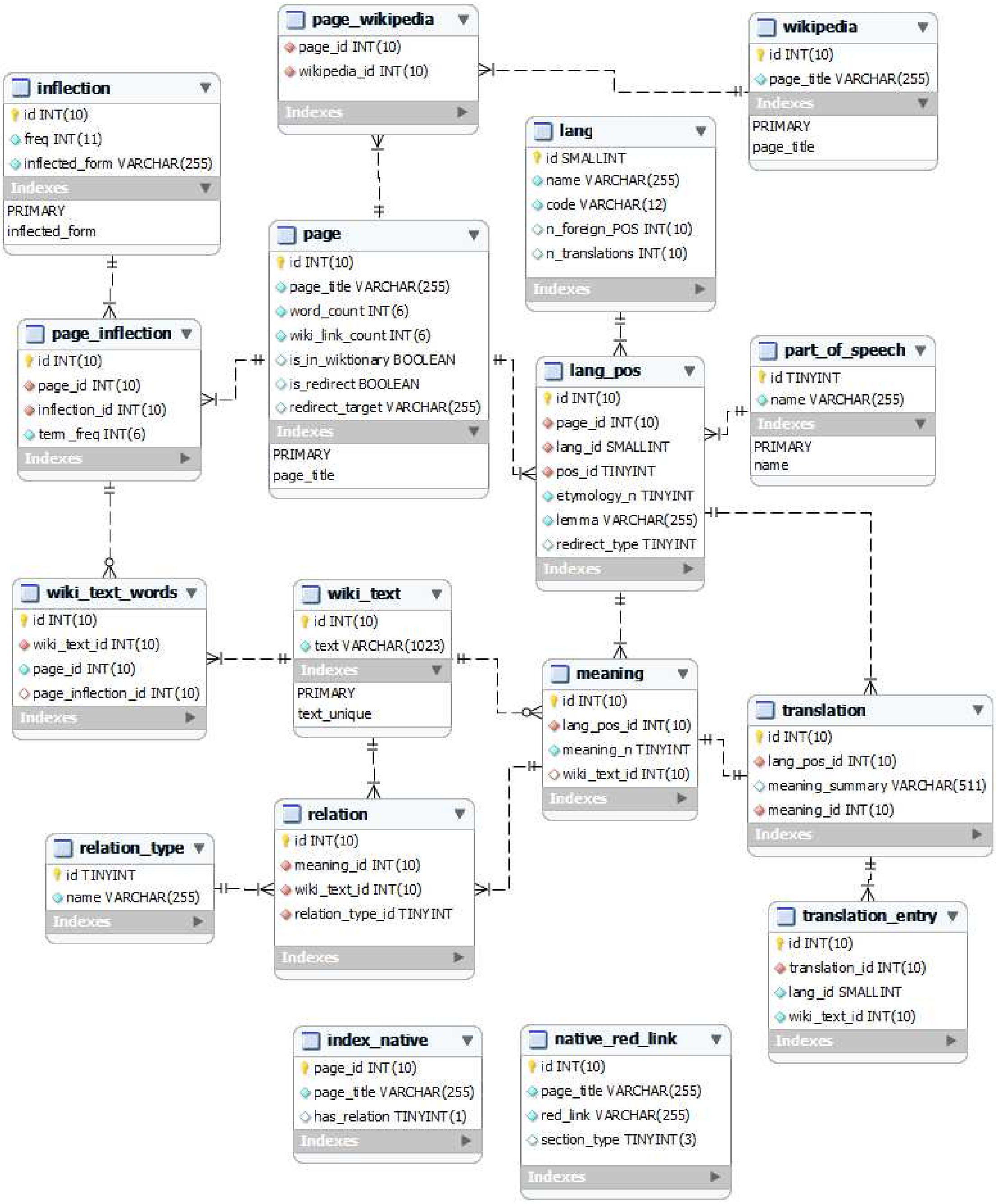

**Figure 6. Tables and relations in the database of the machine-readable dictionary**

# 4. CONCLUSION

The architecture of the extensible and modular Wiktionary parser was developed. The modules for the extraction of three types of data from the Wiktionary entries (i.e. data of three subsections) were implemented: words meanings, semantic relations and translations. These modules were adapted to the English Wiktionary and the Russian Wiktionary, since formatting rules and the structure of the entry are different.

Many times (during the pass) various regular expressions are used to extract the desired information from the text. This extraction and analysis are possible only due to the known structure of the article and due to the applying of the templates.[10] The more strict and rigid structure of the entry is adopted by a community of wiki editors, the more simple and reliable will the parser's algorithms be. The more number of templates widely used in the Wiktionary is, the more easy to extract the structured data from it.

The transforming of an implicit structure (i.e. structure that is understandable for the Wiktionary reader) into the explicit, unambiguous, machine-understandable structure is presented in the paper. That is, the correspondences between fragments (sections) of a Wiktionary entry, on the one hand, and tables in the database of the machine-readable dictionary (required in order to store data extracted from the Wiktionary), on the other hand, are described and discussed. After that automatic transformation, the Wiktionary data could be easily used in projects related to the language processing.

The creation of machine-readable dictionaries is an important step in the road of the automatic text processing. Machine-readable dictionaries and Wikipedias, and Wiktionaries are heavily used in different disciplines, including ontology building [14], machine translation [1], [11], automatic text simplification [12], image search [1], and word sense disambiguation [9].

There are many attractive ways to develop the parser and applications based on it. But, in the first place, the Graphical User Interface for the created machine-readable dictionary based on data from the English Wiktionary should be developed. This interface for the Russian Wiktionary is ready and available online.[11]

[10] See http://en.wiktionary.org/wiki/Wiktionary:Templates

[11] See the program *wiwordik* based on the data extracted from the Russian Wiktionary: http://code.google.com/p/wikokit/wiki/wiwordik

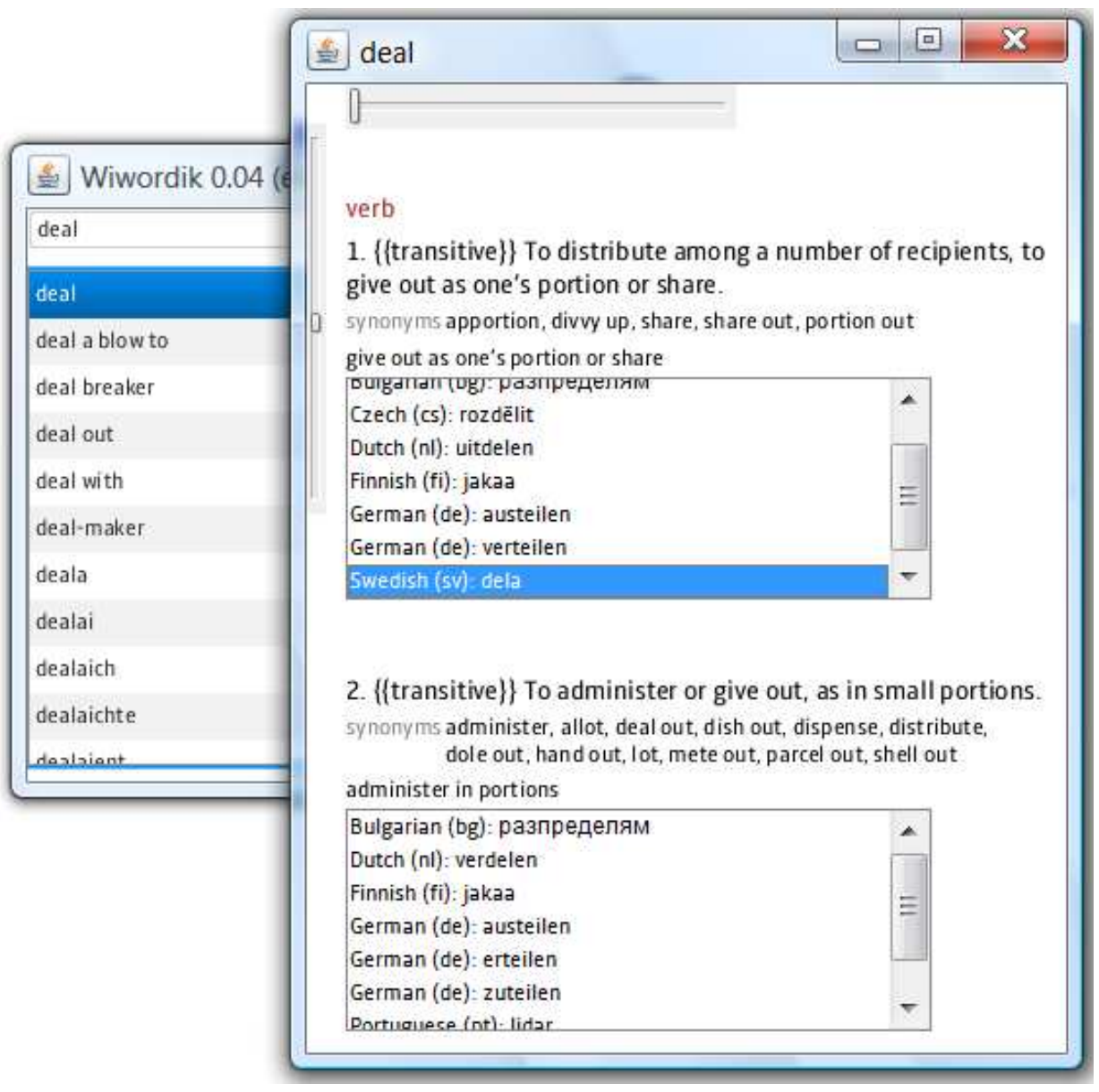

**Figure 7. The word card of the entry *deal* allows visualization of data from the machine-readable dictionary**

# 5. ACKNOWLEDGMENTS

The paper is due to the research carried out as part of projects funded by grants 08-07-00264, 09-07-00436, 09-07-00066, 09-07-12111, and 11-01-00251 of the Russian Foundation for Basic Research, and project 213 of the research program "Intelligent information technologies, mathematical modeling, system analysis and automation" of the Russian Academy of Sciences.